\documentclass[12pt]{nostro}

\usepackage{epsfig,amssymb,amsfonts,wasysym}\parskip 5pt plus 1pt
\newcommand{\be}{\begin{equation}}
\newcommand{\ee}{\end{equation}}
\newcommand{\bea}{\begin{eqnarray}}
\newcommand{\eea}{\end{eqnarray}}

\def\simge{\mathrel{%
   \rlap{\raise 0.511ex \hbox{$>$}}{\lower 0.511ex \hbox{$\sim$}}}}
\def\simle{\mathrel{
   \rlap{\raise 0.511ex \hbox{$<$}}{\lower 0.511ex \hbox{$\sim$}}}}

\begin{document}
%
\begin{frontmatter}
\thispagestyle{empty}
\begin{flushright}
\end{flushright}
\title{The neutrino velocity anomaly as an explanation of the missing observation of neutrinos in coincidence with GRB}
\author[Lione]{D. Autiero}
\author[Napoli]{P. Migliozzi}
\author[Napoli]{A. Russo}
\address[Lione]{IPNL, Universit\'e Claude Bernard Lyon I, CNRS/IN2P3, F-69622 Villeurbanne, France}
\address[Napoli]{I.N.F.N., Sezione di Napoli, Complesso Universitario Monte Sant'Angelo, I-80125, Naples, Italy}
\vspace{.3cm}
\begin{abstract}
The search for neutrinos emitted in coincidence with Gamma-Bay Burst has been so far unsuccessfully. In this paper we show that the recent result
 reported by the OPERA Collaboration on an early arrival time of muon neutrinos with respect to the one computed assuming the speed of light in vacuum
 could explain the null search for neutrinos in coincidence with Gamma-Ray Burst.
\end{abstract}

\vspace*{\stretch{2}}
\begin{flushleft}
  \vskip 2cm
  \small
\end{flushleft}
\end{frontmatter}

\section{Introduction}
\label{sec:intro}
Gamma-Bay Burst (GRB) are among the most violent events in the Universe and could be one of the possible source of ultra-high energy cosmic-rays.
The currently leading model for GRBs production is the fireball model \cite{Waxman:1997ti,Guetta:2003wi}. Schematically, it explains the emission
of GRBs by considering an energy source (central engine) consisting of a newly formed black hole that rapidly increases its mass.  Within the
fireball model framework, observed $\gamma$-rays are produced by synchrotron emission of electrons accelerated to high energy by internal shocks
within the expanding wind. In addition to electrons, protons are also expected to be accelerated through the Fermi mechanism and may interact with
the keV-MeV photons forming a $\Delta^+$ resonance that decays into pions. From the subsequent pion decay a detectable flux of high-energy
neutrinos with energy around $10^{14}$ eV is expected to be emitted in coincidence with GRBs. The production neutrino flavor ratio expected from
this model is ($\nu_e;\nu_\mu;\nu_\tau$)=(1;2;0). However, due to the neutrino oscillation mechanism this ratio is changed into
(1;1;1)\cite{Learned:1994wg,Athar:2005wg}. According to the fireball model a detector of $\sim$ 1 km$^2$ should detect about 10-100 muonic
neutrino interactions per year in coincidence with GRB. On top of the fireball approach, there are other models that describe the GRB production.
It is worth mention that they also demand production of high-energy neutrinos in coincidence with GRBs see \cite{Becker:2009zc,Gao:2011jt} and
references therein.

This Letter is organized as follow: after a brief review of the search for high-energy neutrinos in coincidence with GRBs, we summarize the
recent results published by the OPERA Collaboration \cite{nuvel} that show a neutrino speed  faster than the speed of light with a statistical
significance of 6.0 $\sigma$. Finally, we discuss the impact of the OPERA results on the search for neutrino emission from GRBs.

\section{Present status of the search for neutrinos emission from Gamma-Ray Bursts}
\label{sec:preres}
The IceCube Neutrino Observatory is a neutrino telescope constructed at the Amundsen-Scott South Pole Station in Antarctica\cite{Halzen:2010yj}.
Similar to its predecessor, the Antarctic Muon And Neutrino Detector Array (AMANDA) \cite{Amanda}, IceCube contains thousands of spherical optical
sensors called Digital Optical Modules (DOMs), each with a photomultiplier tube (PMT) and a single board data acquisition computer which sends
digital data to the counting house on the surface above the array  IceCube was completed on the 18th December, 2010 see \cite{icecube}.

The main goal of IceCube is to detect point sources of neutrinos that could help explain the mystery of the origin of the highest energy
cosmic rays. Astrophysical events which are cataclysmic enough to create such high energy particles would probably also create high energy
neutrinos, which could travel to the Earth with very little deflection, because neutrinos interact so rarely. IceCube could observe these
neutrinos: its observable energy range is about 100 GeV to several PeV.

As explained in Section \ref{sec:intro} when protons collide with one another or with photons, the result is usually pions. Charged pions
decay into muons and muon neutrinos whereas neutral pions decay into gamma rays. Potentially, the neutrino flux and the gamma ray flux may
coincide in certain sources such as gamma ray bursts and supernova remnants, indicating the elusive nature of their origin. Therefore, the
IceCube Collaboration performed a dedicated search for neutrino emission from GRB sources.

The search discussed in \cite{Abbasi:2011qc} has been conducted as follow. A catalog of GRB observed in the Northern Emisphere has been assembled via the GRB coordinates
Network (GCN) \cite{GRBNet}. A search for possible neutrino signal has been performed in a time window around the observed GRB. From April 2008 to
May 2009 IceCube was operating with 40 out of the 86 strings composing the whole detector. During this period 129 GRBs were observed
in the Northern Emisphere. Two different analyses have been conducted to exploit this sample of GRBs. One dedicated to the search for neutrinos
produced in $p-\gamma$  interactions in the prompt phase of the GRB fireball (model dependent analysis); a second  more generic search for any
neutrino emission from GRB over a wide range of energies and emission times (model independent analysis). Both searches gave a null results for
neutrino emission in coincidence with GRBs. The IceCube Collaboration in the conclusions of \cite{Abbasi:2011qc} wrote \emph{While the specific
neutrino-flux predictions of the fireball model provided by Waxman and Bahcall \cite{Waxman:1997ti} and by Guetta et al. \cite{Guetta:2003wi} are
excluded (90\% confidence) by this work, we have not yet ruled out the general picture of fireball phenomenology. The neutrino flux we compute for
GRBs is determined by the flux of protons accelerated in the fireball, and by the fraction of proton energy transferred to charged pions ($f_\pi$).
The proton flux can be chosen either such that the energy in gammas and protons is equal or set to the flux of cosmic rays above $10^{18}$ eV, with
similar results. $f_\pi$ is determined largely by assuming protons are accelerated, in conjunction with the observed low optical thickness of the
source.
Due to uncertainties in the bulk boost factor and internal structure of the shocks, $f_\pi$ may range from $10 - 30\%$ \cite{Guetta:2001cd},
causing an uncertainty of about a factor of 2 on our calculation of the flux, which used  $f_\pi\sim0.2$. Future observations by IceCube will push
our sensitivity below the level of this theoretical uncertainty on $f_\pi$ and allow direct constraints on acceleration of protons to ultra-high
energies in Gamma Ray Bursts.} Recently preliminary results from a search with 59 out of the 86 strings composing the whole detector has been
presented \cite{icrc2011}. No neutrinos have been observed in coincidence with a GRB, while 8 events were expected.

\section{Impact of neutrino velocity measurement on GRB neutrino searches}
\label{sec:results}
The OPERA neutrino experiment \cite{operafirst} in the underground Gran Sasso Laboratory (LNGS) was designed to perform the first detection of neutrino
oscillations in direct appearance mode in the $\nu_\mu\rightarrow\nu_\tau$ channel channel, the $\nu_\tau$ signature being the identification
of the $\tau$ lepton created in its charged current (CC) interaction \cite{operacandidate}.
A precision measurement of the neutrino velocity by using muonic neutrinos from the CNGS beam \cite{CNGS} has been reported in [REF].
This measurement was made possible thanks to the high-statistics data taken by the OPERA detector, by dedicated upgrades of the CERN
timing system for the time tagging of the CNGS beam and of the OPERA detector resulting in a reduction of the systematic uncertainties down
to the level of the statistical error ($\sim$ 10 ns). Finally, by a high-accuracy geodesy campaign that allowed measuring the 730 km CNGS baseline
with a precision of the order of 20 cm. All the details are given in \cite{nuvel} and references therein.

Having defined $\delta t$ as the difference between $TOF_c$ (the expected time of flight from the neutrino source to the detector assuming
the speed of light) and  $TOF_\nu$ (the time of flight of CNGS neutrinos from the neutrino source to the detector),
i.e. $\delta t = TOF_c - TOF_\nu$, the final result on the measurement is

$$\delta t = (60.7 \pm 6.9 (stat.) \pm 7.4 (sys.))\,\mbox{ns.}$$

Namely, the measurement indicates an early arrival time of CNGS muon neutrinos with respect to a light signal with a 6$\sigma$ significance.

How this result affects the search for neutrinos in coincidence with GRBs? Here, we do not want to attempt theoretical interpretations of the result
on neutrino velocity, i.e. is there an energy dependence of the neutrino velocity? Therefore, let's assume conservatively that the
anticipation measured by OPERA is constant also at higher neutrino energies.

Typically, the distance from where a GRB originated is measured in terms of the quantity $z$, i.e. the source read-shift, and $z=1$ corresponds
to a distance of $1.302773\times10^{+23}$ km. In the sample of GRBs analyzed by the IceCube Collaboration the average $z$ is about 2 and ranges
between 0.8 and 2.5 \cite{Abbasi:2011qc}.

If we assume that every 730 km the neutrinos anticipate by 60 ns the light, i.e. no energy dependence is assumed, we can compute the anticipation
of the neutrinos arrival time with respect to the light as a function of $z$. The results are shown in Table \ref{tab:total}. Even for the closest
GRBs the neutrino arrival time is anticipated, with respect to the light arrival time, by more than hundred thousand years! Therefore, this result
could explain why the specific neutrino-flux predictions of the fireball model provided by Waxman and Bahcall \cite{Waxman:1997ti} and by
Guetta et al. \cite{Guetta:2003wi} are excluded (90\% confidence) by the IceCube experiment. Nevertheless, the general picture of the fireball
phenomenology is not yet ruled out.

The forthcoming results of the IceCube Collaboration on the search for neutrinos from GRB are, therefore, of the outmost importance. Indeed, a
null result would further exclude the neutrino-flux predictions of the fireball model. Such a result could then be explained either by a complete
failure of the fireball model or by advocating new phenomena. Unfortunately, neutrino telescopes will be never able to measure an earlier
time arrival of neutrinos  from GRB, but they could only give indirect support to the OPERA result. On the other hand, an unambiguous
detection of neutrino from GRBs would  disprove the interpretation of the OPERA result in terms of a neutrino traveling faster than light.

\begin{table}[!h]
\begin{center}
\begin{tabular}{|c|c|c|c|}
\hline
$z$ & Mpc & Neutrino anticipation ($10^{10}$ s) & Neutrino anticipation ($10^{5}$ year) \\
\hline
0,5 & 1657,67 & 417,73  & 2,65\\
\hline
1   & 3315,33 & 835,46  & 4,22\\
\hline
2   & 5275,05 & 1329,31 & 5,90\\
\hline
4   &h 7383,55 & 1860,65 & 7,41\\
\hline
\end{tabular}
\caption{\small{ Anticipation of the neutrinos arrival time with respect to the light as a function of $z$.}}
\label{tab:total}
\end{center}
\end{table}





\begin{thebibliography}{999}


\bibitem{Waxman:1997ti}
  E.~Waxman, J.~N.~Bahcall,
  Phys.\ Rev.\ Lett.\  {\bf 78 } (1997)  2292-2295.
  [astro-ph/9701231].

\bibitem{Guetta:2003wi}
  D.~Guetta, D.~Hooper, J.~Alvarez-Muniz, F.~Halzen, E.~Reuveni,
  Astropart.\ Phys.\  {\bf 20 } (2004)  429-455.
  [astro-ph/0302524].


\bibitem{Learned:1994wg}
  J.~G.~Learned, S.~Pakvasa,
  Astropart.\ Phys.\  {\bf 3 } (1995)  267-274.
  [hep-ph/9405296, hep-ph/9408296].

\bibitem{Athar:2005wg}
  H.~Athar, C.~S.~Kim, J.~Lee,
  Mod.\ Phys.\ Lett.\  {\bf A21 } (2006)  1049-1066.
  [hep-ph/0505017].


\bibitem{Becker:2009zc}
  J.~K.~Becker, F.~Halzen, A.~O Murchadha, M.~Olivo,
  Astrophys.\ J.\  {\bf 721 } (2010)  1891-1899.
  [arXiv:0911.2202 [astro-ph.HE]].

\bibitem{Gao:2011jt}
  S.~Gao, K.~Toma, P.~Meszaros,
  Phys.\ Rev.\  {\bf D83 } (2011)  103004.
  [arXiv:1103.5477 [astro-ph.HE]].

\bibitem{nuvel}
  T.~Adam et al., ``Measurement of the neutrino velocity with the OPERA detector in the CNGS beam ,'' [arXiv:1109.4897 [hep-ex]].

\bibitem{Halzen:2010yj}
  F.~Halzen, S.~R.~Klein,
  Rev.\ Sci.\ Instrum.\  {\bf 81 } (2010)  081101.
  [arXiv:1007.1247 [astro-ph.HE]].


\bibitem{Amanda}
  E.~Andr\'es et al. [AMANDA Collaboration],
  Nature\  {\bf 410 } (2001)  441.

\bibitem{icecube}
IceCube Collaboration, material available at  http://www.icecube.wisc.edu

\bibitem{Abbasi:2011qc}
IceCube Collaboration, R.~Abbasi {\it et al.},
  Phys.\ Rev.\ Lett.\  {\bf 106 } (2011)  141101.
  [arXiv:1101.1448 [astro-ph.HE]].

\bibitem{GRBNet}
"GRB	Coordinates Netweork", http://gcn.gsfc.nasa.gov/.

\bibitem{Guetta:2001cd}
  D.~Guetta, M.~Spada, E.~Waxman,
  Astrophys.\ J.\  {\bf 559 } (2001)  101.
  [astro-ph/0102487].

\bibitem{icrc2011}
IceCube Collaboration, H. Kolanoski, talk given at ICRC 2011, Bejing, China.

\bibitem{operafirst}
OPERA Collaboration, R. Acquafredda et al., New J. Phys. 8 (2006) 303;\\
OPERA Collaboration, R. Acquafredda et al., JINST 4 (2009) P04018;\\
OPERA Collaboration, N. Agafonova et al.,JINST 4 (2009) P06020.

\bibitem{operacandidate}
OPERA Collaboration, N. Agafonova et al., Phys. Lett. B 691 (2010) 138.

\bibitem{CNGS}
Ed. K. Elsener,  ÒThe CERN Neutrino beam to Gran Sasso (Conceptual Technical Design)Ó, CERN 98-02, INFN/AE-98/05;\\
R. Bailey et al., ÒThe CERN Neutrino beam to Gran Sasso (NGS)Ó (Addendum to report CERN 98-02, INFN/AE-98/05)Ó, CERN-SL/99-034(DI), INFN/AE-99/05.

\end{thebibliography}
\end{document}